# Enabling variable high spatial resolution retrieval from a long pulse BOTDA sensor

Zhao Ge, Li Shen, Can Zhao, Hao Wu, Zhiyong Zhao, and Ming Tang, *Senior Member, IEEE*

*Abstract*—In the field of Internet of Things, there is an urgent need for sensors with large-scale sensing capability for scenarios such as intelligent monitoring of production lines and urban infrastructure. Brillouin optical time domain analysis (BOTDA) sensors, which can monitor thousands of continuous points simultaneously, show great advantages in these applications. We propose a convolutional neural network (CNN) to process the data of conventional Brillouin optical time domain analysis (BOTDA) sensors, which achieves unprecedented performance improvement that allows to directly retrieve higher spatial resolution (SR) from the sensing system that use long pump pulses. By using the simulated Brillouin gain spectrums (BGSs) as the CNN input and the corresponding high SR BFS as the output target, the trained CNN is able to obtain a SR higher than the theoretical value determined by the pump pulse width. In the experiment, the CNN accurately retrieves 0.5-m hotspots from the measured BGS with pump pulses from 20 to 50 ns, and the acquired BFS is in great agreement with 45/40 ns differential pulse-width pair (DPP) measurement results. Compared with the DPP technique, the proposed CNN demonstrates a 2-fold improvement in BFS uncertainty with only half the measurement time. In addition, by changing the training datasets, the proposed CNN can obtain tunable high SR retrieval based on conventional BOTDA sensors that use long pulses without any requirement of hardware modifications. The proposed data post-processing approach paves the way to enable novel high spatial resolution BOTDA sensors, which brings substantial improvement over the state-of-the-art techniques in terms of system complexity, measurement time and reliability, etc.

## I. Introduction

With the development of smart factories, smart cities, industrial Internet of Things (IoT), etc., the IoT has dramatically changed our lives. Brillouin optical time domain analysis (BOTDA) can measure the temperature and strain distribution along the optical fibers with tens of kilometers range[1-6], so it is naturally suitable for large-scale and high-density sensing monitoring of IoT, such as production line monitoring, urban infrastructure monitoring, etc. Spatial resolution (SR) is one of the most important parameters of a BOTDA sensor[7-11], which represents the minimum fiber length required to accurately measure the sensing information. Conventionally, the SR is determined by the pump pulse width, and is normally restricted to be longer than 1 m due to the ~10 ns acoustic lifetime limitation in silica optical fibers. Various approaches have been proposed to enable sub-meter SR, one of the most commonly used methods is the differential pulse-width pair (DPP) technique[12-16]. In DPP-BOTDA, two Brillouin time-domain traces are separately measured using two long pump pulses with a width difference, and high SR sensing signals can be obtained by subtracting the measured traces from each other. DPP-BOTDA has the advantages of easy implementation and long sensing distance, but it takes twice the measurement time and are more vulnerable to polarization fading noise and system instability[17]. Due to these drawbacks, it's necessary to explore new ways to obtain high SR directly from the measured Brillouin gain spectrum (BGS) with long pump pulses, which can reduce the measurement time while improving the Brillouin frequency shift (BFS) accuracy.

Recently, signal post-processing methods have been proposed to achieve this purpose[18-20], which can significantly improve the SR by analyzing the BGS with specially designed algorithms. In pump pulse subdivision method, the measurement results of a long pump pulse are considered as a superposition of several short pulses, and high SR can be retrieved by recovering the sensing signals related to the short pulse[20]. Another kind of method to improve SR is to use deconvolution algorithm. By approximating the Brillouin time-domain traces as a linear convolution between pump pulse shape and fiber impulse response, deconvolution algorithm successfully obtains a 0.2-m SR from the measured BGS with 40 ns pump pulse[17]. Post-processing methods break the SR limitation resulting from pump pulse width, and can achieve flexible adjustment of SR. However, because of the inertial features of acoustic wave, BOTDA sensors cannot be rigorously regarded as a linear time-invariant system[17, 21]. Consequently, the Brillouin gain envelope is influenced by the detuned frequency along the fiber, and the above-mentioned post-processing methods will lead to notable distortions in the recovered results where BFS has a sharp and large change[17].

This work was supported in part by National Key Research and Development Program of China (2018YFB1801002), in part by National Natural Science Foundation of China (61931010), in part by Fundamental Research Funds for the Central Universities (HUST: 2021XXJS026), in part by Hubei Province Key Research and Development Program (2020BAA006), in part by Innovation Fund of WNLO. (Zhao Ge and Li Shen contributed equally to this work.) (Corresponding author: Hao Wu; Zhiyong Zhao.)

The authors are with the School of Optical and Electronic Information, Huazhong University of Science and Technology, Wuhan 430074, China (e-mail: m201977113@hust.edu.cn; shenli@hust.edu.cn; zhao_can@hust.edu.cn; wuhaoboom@qq.com; zhiyongzhao@hust.edu.cn; tangming@mail.hust.edu.cn).

Although some sophisticated pre-processing methods have been proposed to eliminate this distortion, it will cause an increasement of measurement time[21].

With the development of artificial intelligence, neural networks are now increasingly applied in BOTDA sensors. Experimental results have demonstrated that for BFS determination, convolutional neural network (CNN) outperforms the traditional Lorentz curve fitting (LCF) method in both precision and computation speed[2]. These results show that CNN has a two-dimensional feature extraction ability and non-linear mapping function, which can comprehensively analyze the features of BGS over a certain length of fiber to acquire the BFS distribution. Therefore, it is expected that the outstanding performance of CNN might also find application in enabling high SR measurement in BOTDA sensors, where CNN may have the capacity to directly achieve high SR by matching the BGS features to the BFS distribution, and avoid distortion caused by conventional post-processing methods.

In this work, a deep learning algorithm based on CNN is proposed as a post-processing method to improve the SR of conventional BOTDA sensors. To train the CNN, a large number of BGSs is simulated as the CNN input training data, and the corresponding high SR BFS distribution is used as the CNN output target. After the training, simulation results indicate that the CNN can realize precise BFS extraction with high SR from the input BGS that is generated by a long pulse. In the experiment, the CNN can accurately recover the 0.5 m SR BFS distribution when 20 to 50 ns pump pulses are used, and the obtained BFS is in great agreement with the 45/40 ns DPP measurement results. By changing the simulated training data sets, variable high SR sensing information can be recovered directly from measured BGS with different long pump pulse widths.

## II. THE ARCHITECTURE OF THE PROPOSED CNN

As shown in Fig. 1, the architecture of the proposed CNN is based on ResNet, which consists of three parts[22]. The first part mainly includes an input layer, a convolutional (Conv) layer with 64 filters of size 7×7, and a maximum pooling layer with 64 filters of size 3×3. The input layer is used for the import of BGS. The size of input data is 71×540, where the height 71 is the number of swept frequencies of the BGS, and the width 540 is the number of BGS along the fiber length direction. By carefully designing the stride length and padding size of the convolutional and pooling layers, as marked by the red numbers in Fig. 1, the size of the 64 output feature maps after the first part is reduced to 18×540, which can greatly reduce the number of unnecessary parameters in the CNN. To alleviate the internal covariate shift problem and add nonlinear factors, batch-normalization (BN) layer and rectified linear unit (ReLU) activation function are also contained in the first part[23,24]

The second part of the CNN is a deep neural network to extract the features of the BGS, which is composed of 16 residual blocks to have a large enough receptive field covering the BGS with length longer than the pump pulse width[22, 25]. Each residual block has two convolutional layers, BN layers, and ReLU activation function, and there is a shortcut connection between the input and output to solve the vanishing gradients and degradation problems of deep neural networks[22]. According to the number of filters in the convolutional layers, the 16 residual blocks can be divided into four categories with 64, 128, 256, and 512 filters, and the numbers of residual blocks in each category are 3, 4, 6, and 3 respectively. As indicated by the red numbers in Fig. 1, during the CNN processing, the size of output feature maps after each network section gradually decreases in the frequency direction, but remains unchanged in the fiber length direction.

The third part of the CNN is a single convolutional layer aims to traverse the data to the one-dimensional BFS distribution. The final output size of the CNN is 1×540, which is the BFS corresponding to each input BGS, thereby accomplish the BFS extraction task.

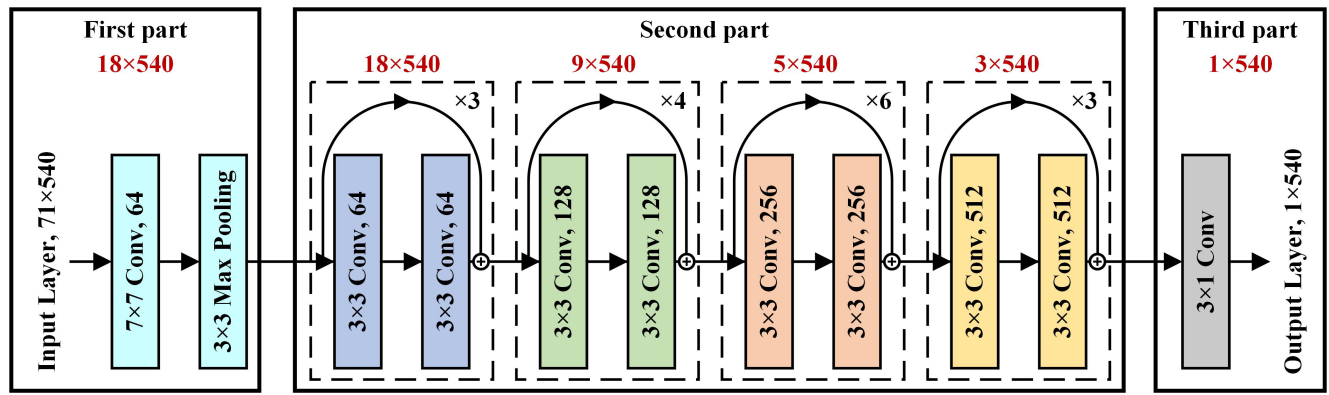

Fig. 1. Architecture of the proposed CNN

## III. SIMULATION RESULTS

### A. Simulation data generation

To make the CNN have an accurate BFS determination ability for various application situations, a large amount of training data is required, which are all generated by mathematical model in this work. In the simulation, the BGS of a 54-m fiber is calculated, which is composed of multiple uniform fiber sections with length from 0.5-5 m. As listed in Table 1, the BFS, section length, normalized gain intensity and intrinsic Brillouin linewidth of each uniform section are set with random distributed values, thus the obtained BGS contains various situations. It's presumed that the shortest length of uniform fiber sections has an important impact on the CNN, because it represents the minimum length of BFS change that the CNN can be applied to after the training, i.e. the SR of the system. According to the uniform fiber sections ranges, the expected SR of the trained CNN is 0.5 m. Considering that the Brillouin gain intensity is affected by temperature and strain change, the normalized gain intensity fluctuates between 0.8-1. It's worth mentioning that the linewidth in Table 1 is the intrinsic Brillouin linewidth, and the simulated BGS linewidth is also related to the pump pulse width, which will be broader when narrow pump pulse is used[26].

With the given simulation parameters, the BGS at fiber position $z$ can be solved by the concatenation of Brillouin gain of lots of very short fiber unit within the pump pulse width, which can be calculated by[27, 28]:

TABLE I
RANDOM RANGES OF SIMULATION PARAMETERS

| Parameters | Random Range |
| --- | --- |
| BFS | 10.81-10.89 GHz |
| Length of Fiber Sections | 0.5-5 m |
| Normalized Gain Intensity | 0.8-1 |
| Intrinsic Brillouin Linewidth | 25-35 MHz |

$$a_s(z) = \sum_{N=0}^{l/\Delta z} a_s^{short}\left(z - N\Delta z, \frac{z + N\Delta z}{V_g}\right) \quad (1)$$

where $l$ is the length of pump pulse, $\Delta z$ is the length of a short fiber unit, which is 1 cm in the simulation, and $a_s^{short}$ is the induced Brillouin gain of one fiber unit, which can be expressed as[27, 28]:

$$a_s^{short}(z,t) = g(z)\frac{I_P^0 A_S^0}{2\Gamma_A^*}\Delta z \left\{1 - \exp\left[-\Gamma_A^*\left(t - \frac{z + \Delta z}{V_g}\right)\right]\right\} \\ \left[u\left(t - \frac{z + \Delta z}{V_g}\right) - u\left(t - T - \frac{z + \Delta z}{V_g}\right)\right] \quad (2)$$

Where $a_s^{short}(z,t)$ is the Brillouin gain at position $z$ and time $t$, $g(z)$ is a constant related to the electrostrictive constant, $I_P^0$ and $A_S^0$ are the pump pulse and constant acoustic wave intensity respectively, $V_g$ is the light velocity in the fiber, $u$ is the Heaviside unit step function, $T$ is the pump pulse width, and $\Gamma_A = i\pi(v_B^2(z) - v^2 - iv\Delta v_B)$ is the frequency detuning parameter, where $v_B$ and $v$ are the BFS at position $z$ and the sweep frequency respectively, and $\Delta v_B$ is the intrinsic Brillouin linewidth[27]. The BGS is simulated with a rectangular 40-ns pump pulse and sampling rate of 1 GSa/s. The frequency sweep range is 10.78-10.92 GHz with a step of 2 MHz to obtain a total of 71 frequencies. Then the obtained BGS is divided into size of 71×540 and normalized according to the maximum Brillouin gain. Finally, Gaussian white noise is added to the normalized BGS with random variance between 0.005-0.0005, resulting in a SNR between 23-33 dB.

On the other hand, as the CNN output target, the corresponding BFS label also needs to be normalized by:

$$BFS_N = \frac{BFS - BFS_{min}}{BFS_{max} - BFS_{min}} \quad (3)$$

where $BFS_N$ is the BFS after normalization, which varies between 0 and 1, $BFS_{max}$ and $BFS_{min}$ are the maximum and minimum values of the BFS range, respectively. Meanwhile, to make the BFS label has a 0.5 m SR as well, a Gaussian filter is applied to the BFS to generate smooth rising/falling edges. It's very important that the bandwidth of the Gaussian filter should not degraded the SR worse than 0.5 m. So, a Gaussian filter with 200 MHz bandwidth is used in the simulation, which corresponds to a SR of 0.5 m exactly[29, 30]. It should be noted that although filtered BFS label allows the CNN to obtain a more stable prediction at BFS changing regions, the rising/falling length will be longer.

B. *Simulation verification of the CNN*

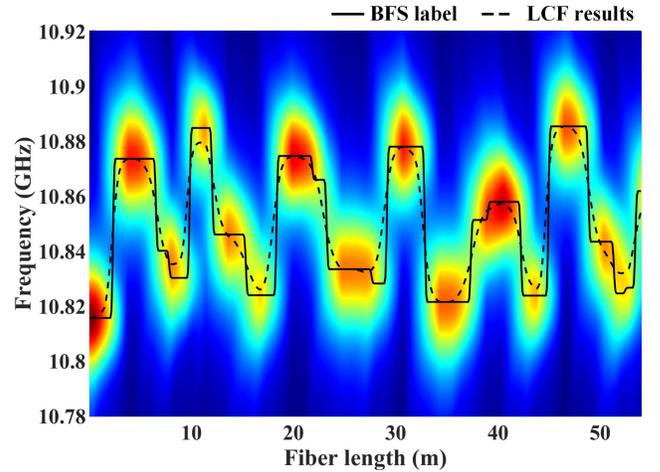

Fig. 2. The simulated BGS and the corresponding BFS label of a set of the training data.

10000 sets of training data are generated, one of which is shown in Fig. 2, where the solid line is the BFS label. We can observe that since the pump pulse width in the simulation is 40 ns, the BGS of uniform fiber sections that is shorter than 4 m will be distorted due to SR limitation. As shown by the dash line of LCF results, accurate BFS cannot be acquired for these distorted regions. Different from the principle of LCF which only analyzes the BGS at single fiber location, with the multi-layer convolutional structure, the predicted BFS of CNN at every location is based on the features of BGS in the receptive field. Although the original SR of input BGS is limited by pump pulse width, during the training process, the CNN is automatically learned to recover a BFS distribution that best fits the output label, which has a SR of 0.5 m. Thus, the trained CNN can achieve higher SR from a measured BGS with long pump pulses.

During the training, the CNN are firstly initialized with Kaiming method, then Adam optimization is employed to minimize the mean squared error (MSE) loss function with a learning rate of $10^{-4}$. Because the BGS on the left and right edges of the input data is affected by the BFS outside the input range, only the output results with center size of 1×500 are used for the calculation of MSE loss. After training the CNN for 82 epochs, the optimal model is obtained with an MSE loss of $9.93×10^{-5}$. It requires about 7.7 hours to complete the training with a NVIDIA TITAN RTX GPU.

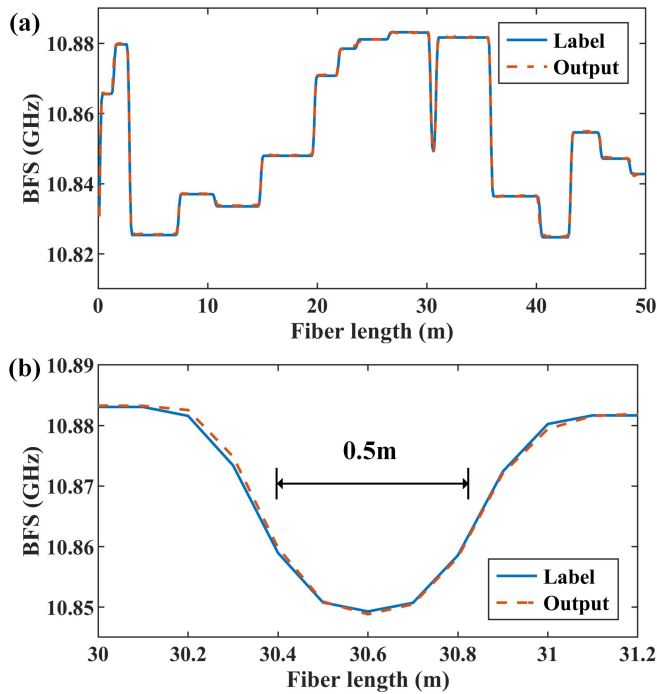

Fig. 3. (a) The comparation between BFS label and CNN output results, (b)The BFS label and CNN output results at 0.5 m changing region.

To evaluate the performance of the trained CNN, a test dataset is generated in the same way as the training dataset. The blue line and red dash line in Fig. 3(a) show the comparison of BFS label and CNN output respectively. It can be discovered that there is no apparent error, and the output results are consistent with the label. For the BFS changing regions with 0.5 m length, as shown in Fig. 3(b), the CNN can accurately extract the BFS distribution without obvious distortion, which proves that the CNN has a SR of 0.5 m. In addition, the SR of this CNN can be flexibly adjusted by changing the training dataset with different target SR.

## IV. EXPERIMENTAL SETUP AND RESULTS

To further verify the effect of the CNN, a typical BOTDA sensor as shown in Fig.4 is used to measure the experimental data. The continuous wave light output from the laser source is divided into probe and pump light by a 50:50 optical coupler. The probe light on the upper branch is modulated by an electro-optical modulator (EOM) to sweep the frequency, which is driven by a radio frequency (RF) generator through carrier suppressed double-sideband modulation. And the sweep range is 10.81 GHz to 10.89 GHz in 2 MHz steps. The probe light is finally launched into the 4.9 km long sensing fiber through an isolator. There are three hotspots placed at the end of the fiber, with lengths of 3.3 m, 1 m and 0.5 m, respectively. The lower branch is used for pump light which is modulated by another EOM to generate high extinction ratio pump pulse with fast rising/falling time by using a programmable electrical pulse generator. The erbium-doped fiber amplifier (EDFA) is used to amplify the pump pulse light, and then the amplified pump pulse light passes through a polarization switch (PS) to reduce the polarization fading noise of the Brillouin gain. At the receiver side, a fiber Bragg grating (FBG) is employed to reflect the Brillouin Stokes sideband. The Brillouin signal is finally detected by a photodetector (PD) and then acquired and displayed on an oscilloscope.

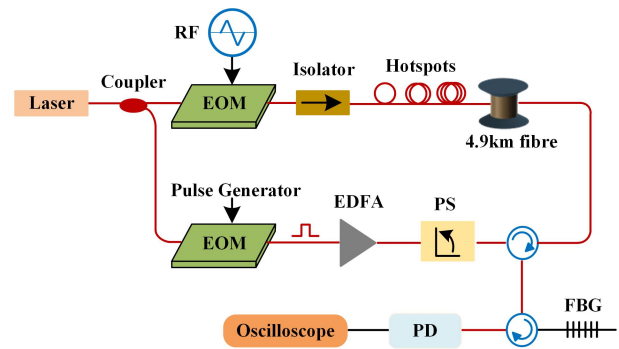

Fig. 4. Experimental setup of the BOTDA system. RF: radio frequency, EOM: electro-optic modulator, PS: polarization switch, EDFA: erbium-doped fiber amplifier, FBG: fiber Bragg grating, PD: photodetector.

A 40 ns pump pulse is used in the experiment as in the simulation, and the sampling rate is 1 GSa/s. The SNR of the experimental raw data is about 25.8 dB. As shown by the blue line in Fig. 5, LCF is used for the obtained BGS, which has a theoretical SR of 4 m. We can discover that the hotspots cannot be accurately measured. Then the experimental data is processed by the trained CNN, and the output BFS distribution near the hotspots is shown as the green dash line. Finally, a 45/40 ns DPP is performed, and the obtained BFS distribution is shown by the red line as a reference. The comparation results indicate that the CNN can accurately retrieve the BFS of the hotspots, and the recovered result is in great agreement with the 45/40 ns DPP.

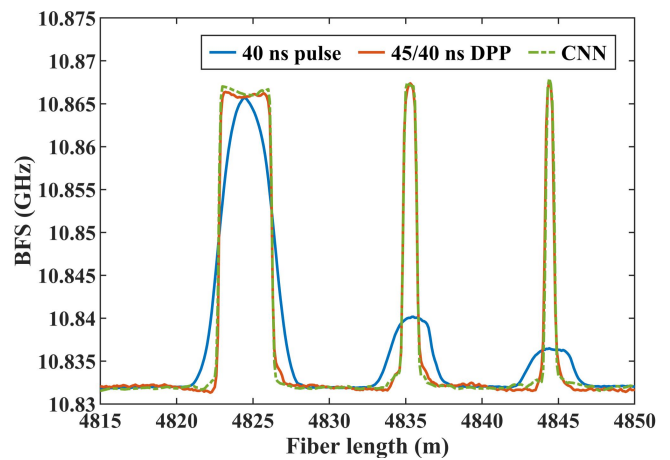

Fig. 5. 40 ns pump pulse measurement results, the CNN output results and 45/40 ns DPP measurement results in the experiment

BFS uncertainty is then compared to quantify the performance of the CNN. In the experiment, the 45/40 ns DPP is measured for 6 times, and the 40 ns pump pulse measurement results are processed by the CNN. The calculated BFS standard deviation of the 6 measurements is regard as BFS uncertainty, which is shown in Fig. 6. The uncertainty and quadratic fitting results of the 45/40 ns DPP is shown by the blue and yellow lines respectively, and the uncertainty and quadratic fitting results of the CNN prediction is shown by the red and purple lines respectively. Under the same SR of 0.5 m, the average uncertainty of CNN prediction along the fiber is 0.31 MHz, which is only half of the 45/40 ns DPP. In addition, compared with the DPP technique, by avoiding the signal differential process, the CNN reduces the required measurement time by half. Also, based on the python environment on the same computer, CNN only needs about 0.14 s to process 10000 BGS, while the LCF used by DPP needs 7.3 s. These results indicate that the proposed high SR BFS retrieval method with CNN has the advantages of short measurement time and fast processing speed.

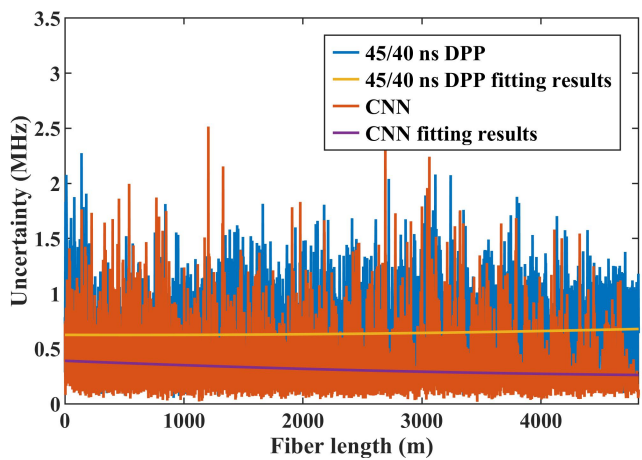

Fig. 6. BFS uncertainty comparation between DPP and CNN for the experimental results

Eventually, to make the CNN applicable to various pump pulse widths, training datasets are also simulated with 20, 30, and 50 ns pump pulses respectively to generate multiple CNNs, and the performance of the CNNs are verified using the experimental data measured with corresponding pump pulse widths. The CNNs processed results are all shown in Fig. 7. For pump pulse widths between 20-50 ns, the obtained BFSs have a good consistency, and the average uncertainty between different pump pulse widths is 0.63 MHz. These experimental results indicate that the CNN can be applied to different pump pulse widths by using matched training data. In addition, it is worth mentioning that by changing the training datasets with different targeted SR, the trained CNNs can achieve variable high SR retrieval from a long pulse BOTDA sensor. The unprecedented adjustable capability of SR that is enabled in software shows much better flexibility than the traditional approaches that requires hardware modification.

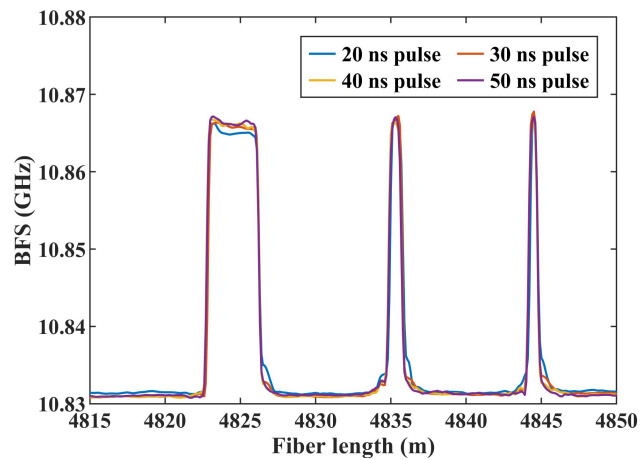

Fig. 7. BFSs retrieved by CNNs from different pump pulse widths between 20-50 ns

## V. Conclusions

In conclusion, in this work a CNN is proposed to directly retrieve the accurate BFS distribution with much higher SR from the measured BGS that is acquired by using a long pump pulse. For proof of concept, in the experiment, 0.5 m SR are successfully obtained with long pump pulses from 20 to 50 ns, and the CNN processed results are in good agreement with the 45/40 ns DPP measurement result, which verifies the excellent feasibility of the proposed data post-processing approach. Under the same SR, the proposed CNN shows a 2-fold improvement in BFS uncertainty over the DPP technique, and consumes only half the measurement time. By designing the training datasets, the proposed CNN with variable SR can be widely used in traditional BOTDA systems to improve the SR without requirement of hardware modification. The proposed technique in this work paves the way to enable a novel high spatial resolution measurement method for BOTDA sensors, which shows unprecedented improvement over the state-of-the-art techniques in terms of system complexity, measurement time and reliability, etc.

**Ge Zhao** received the B.S. degree from the School of Physics and Information Engineering, Jianghan University, Wuhan, China, in 2019.

He is currently working toward the M.D. degree with the School of Optical and Electronic Information, HUST. His research interests are optical fiber sensing and optical fiber devices.

**Li Shen** received the B.S. degree from the School of Optical and Electronic Information, Huazhong University of Science and Technology (HUST), Wuhan, China, in 2016.

He is currently working toward the Ph.D. degree with the School of Optical and Electronic Information, HUST. His research interests are optical fiber sensing and optical fiber devices.

**Can Zhao** received the B.S. and Ph.D. degrees from the School of Optical and Electronic Information, Huazhong University of Science and Technology (HUST), Wuhan, China, in 2014 and 2019, respectively.

He has been working as a postdoctoral researcher at HUST since 2019. His current research interests include distributed optical fiber sensing and specialty optical fiber.

**Hao Wu** received the B.S., M. Eng., and Ph. D. degrees from the School of Optical and Electronic Information, Huazhong University of Science and Technology (HUST), Wuhan, China, in 2013, 2016 and 2019, respectively.

He has been working as a postdoctoral researcher at HUST since 2019. His current research interests include the application of specialty optical fiber and machine learning algorithm in distributed optical fiber sensing.

**Zhiyong Zhao** received the B.Eng. and Ph.D. degrees from Huazhong University of Science and Technology, Wuhan, China, in 2012 and 2017, respectively.

He was a joint Ph.D. Student with the École polytechnique fédérale de Lausanne, Lausanne, Switzerland, from October 2014 to October 2015. In 2016, He was a Research Assistant with the School of Electrical and Electronic Engineering, Nanyang Technological University (NTU), Singapore. From June 2017 to October 2020, he was a Postdoctoral Fellow with the Department of Electronic and Information Engineering, Hong Kong Polytechnic University, Hong Kong. Since October 2020, he has been an Associate Professor with School of Optical and Electronic Information, Huazhong University of Science and Technology, Wuhan, China. His current research interests include optical fiber sensing, optical fiber devices, special optical fibers, and nonlinear fiber optics.

**Ming Tang** (SM'11) received the B.E. degree from Huazhong University of Science and Technology (HUST), Wuhan, China, in 2001, and the Ph.D. degree from Nanyang Technological University, Singapore, in 2005.

His Postdoctoral Research in the Network Technology Research Centre (NTRC) was focused on optical fiber amplifiers, high-power fiber lasers, nonlinear fiber optics, and all-optical signal processing. From February 2009, he was with the Tera-photonics group led by Prof. Hiromasa Ito in RIKEN, Sendai, Japan, as a Research Scientist conducting research on terahertz-wave generation, detection, and application using nonlinear optical technologies. Since March 2011, he has been a Professor with the School of Optical and Electronic Information, Wuhan National Laboratory for Optoelectronics, HUST, Wuhan, China. His current research interests are concerned with optical fiber based linear and nonlinear effects for communication and sensing applications. He has been a member of the IEEE Photonics Society since 2001.